\documentclass[fleqn,usenatbib]{mnras}

\usepackage{newtxtext,newtxmath}

\usepackage[T1]{fontenc}

\DeclareRobustCommand{\VAN}[3]{#2}
\let\VANthebibliography\thebibliography
\def\thebibliography{\DeclareRobustCommand{\VAN}[3]{##3}\VANthebibliography}

\usepackage{graphicx}	
\usepackage{amsmath}	

\newcommand{\logg}{log~$g$}
\newcommand{\teff}{T$_{\text{eff}}$}
\newcommand{\cannon}[0]{\textsc{The Cannon}}
\newcommand{\tess}[0]{\textit{TESS}}
\newcommand{\ktwo}[0]{\textit{K2}}
\newcommand{\dnu}[0]{$\Delta\nu$}
\newcommand{\dP}[0]{$\Delta P$}
\newcommand{\Ctwo}[0]{$^{12}$C$^{14}$N}
\newcommand{\Cthree}[0]{$^{13}$C$^{14}$N}
\newcommand{\Cratio}[0]{$^{12}\rm{C}/^{13}\rm{C}$}

\title[Carbon tracks red giant evolution]{The role of carbon in red giant spectro-seismology\thanks{Based on observations collected with the high-resolution ($R\sim80,000$) {\it Veloce} spectrograph at the Anglo-Australian Telescope.}}

\author[K. A. Banks et al.]{
Kirsten A. Banks,$^{1,2}$\thanks{E-mail: k.banks@unsw.edu.au}
Sarah L. Martell,$^{1,2,3}$
C. G. Tinney,$^{1}$
Dennis Stello,$^{1,2,4}$
Marc Hon$^{5}$,
Claudia Reyes,$^{1}$
\newauthor James Priest,$^{1}$
Sven Buder,$^{6,2}$
and Benjamin T. Montet$^{1,3}$
\\
$^{1}$School of Physics, University of New South Wales, Sydney, NSW 2052, Australia\\
$^{2}$ARC Centre of Excellence for All Sky Astrophysics in 3 Dimensions (ASTRO 3D), Australia\\
$^{3}$UNSW Data Science Hub, University of New South Wales, Sydney, NSW 2052, Australia\\
$^{4}$Sydney Institute for Astronomy (SIfA), School of Physics, University of Sydney, NSW 2006, Australia\\
$^{5}$Institute for Astronomy, University of Hawai`i, 2680 Woodlawn Drive, Honolulu, HI 96822, USA\\
$^{6}$Research School of Astronomy and Astrophysics, Australian National University, Canberra, ACT 2611, Australia}

\date{Accepted XXX. Received YYY; in original form ZZZ}

\pubyear{2024}

\begin{document}
\label{firstpage}
\pagerange{\pageref{firstpage}--\pageref{lastpage}}
\maketitle

\begin{abstract}
Although red clump stars function as reliable standard candles, their surface characteristics (i.e. \teff, \logg, and [Fe/H]) overlap with those of red giant branch stars, which are not standard candles. Recent results have revealed that spectral features containing carbon (e.g. CN molecular bands) carry information correlating with the ``gold-standard'' asteroseismic classifiers that distinguish red clump from red giant branch stars. However, the underlying astrophysical processes driving the correlation between these spectroscopic and asteroseismic quantities in red giants remain inadequately explored. This study aims to enhance our understanding of this ``spectro-seismic'' effect, by refining the list of key spectral features predicting red giant evolutionary state. In addition, we conduct further investigation into those key spectral features to probe the astrophysical processes driving this connection. We employ the data-driven \cannon\ algorithm to analyse high-resolution ($R\sim80,000$) {\it Veloce Rosso} spectra from the Anglo-Australian Telescope for 301 red giant stars (where asteroseismic classifications from the \tess\ mission are known for 123 of the stars). The results highlight molecular spectroscopic features, particularly those containing carbon (e.g. CN), as the primary indicators of the evolutionary states of red giant stars. Furthermore, by investigating CN isotopic pairs (that is, \Ctwo\ and \Cthree) we find suggestions of statistically significant differences in the reduced equivalent widths of such lines, suggesting that physical processes that change the surface abundances and isotopic ratios in red giant stars, such as deep mixing, are the driving forces of the ``spectro-seismic'' connection of red giants.

\end{abstract}




\section{Introduction} \label{sec:Intro}

Red clump (RC) stars are effective standard candles. 
The RC represents a stage of stellar evolution experienced by all low mass ($0.8-2.0\,M_\odot$) stars. Following the red giant branch (RGB) stage of evolution, where an inert helium core is surrounded by a shell of hydrogen fusion, during the RC phase both core helium burning and shell hydrogen burning are active. The ignition of this core helium fusion happens at the same core mass \citep[$0.47\,M_\odot$;][]{Girardi16} regardless of the original mass of the star. As a result, RC stars possess well-confined luminosities, with $\log(L_{\rm{RC}}/L_\odot)=1.95$, and have a mean dispersion of $\sim0.17\pm0.03$\,mag in all colour bands \citep{Hawkins17}.

One limitation of RC stars as standard candles is that they have very similar surface features, i.e. \teff, \logg, [Fe/H], and luminosity, to lower RGB stars, which are not standard candles. This makes RC stars difficult to classify accurately using photometry and stellar parameters. However, the disambiguation of the RC from the RGB can be achieved effectively using asteroseismology \citep{Montalban10,Bedding11,Chaplin13,Stello13,Mosser14}. 

RC and RGB stars exhibit distinct asteroseismic oscillations due to their different internal structures. At the onset of helium fusion, convection is present in the helium core accompanied by an increase in the size of the core. This decreases the density contrast between the core and the surrounding hydrogen-burning shell \citep{Montalban10,Montalban13}, resulting in significant differences in the Brunt–Väisälä frequency. Therefore, RC and RGB stars exhibit different ranges of large frequency separations, \dnu, and period spacing, \dP, and are as a result easily distinguishable in $\Delta\nu-\Delta P$ space \citep{Chaplin13}. 

The oscillations present in red giants are measurable with sufficiently long time-series observations of photometry or radial velocity \citep[e.g. 82 days for \ktwo;][]{Hon18}. However, it has been shown recently that this asteroseismic information is imprinted into the spectra of RC and RGB stars, beyond the radial velocity signal driven by Solar-like oscillations \citep[e.g.,][]{Hawkins18,Casey19,Banks23}. This allows for the efficient classification of many more RC stars due to the existence of 
spectroscopic survey data 
capturing more stars across a larger volume of the Galaxy \citep[e.g. APOGEE, GALAH;][respectively]{Majewski17,Buder21} compared to asteroseismic surveys \citep[e.g. CoRoT, Kepler, \ktwo, \tess;][respectively]{Baglin06,Gilliland10,Howell14,Ricker15}

\cite{Hawkins18} used the data-driven algorithm \cannon\ \citep{Ness15,Casey16} to determine whether red giant spectra could effectively predict asteroseismic parameters such as the large frequency separation and period spacing (i.e. \dnu\ and \dP), or classify RC and RGB stars. They trained \cannon\ on 1,676 red giants with single epoch infrared spectra from the APOGEE survey \citep[$1.5-1.7\,\mu$; ][]{Majewski17}. They found that the spectra of RC and RGB stars of similar temperatures, surface gravities and metalicities are overall quite similar except around molecular CN and CO line features. \cite{Hawkins18} suggest that the differences in these molecular features are to be expected from mixing along the RGB causing RC stars to have a lower [C/N] ratio compared to RGB stars with similar \teff, \logg, and [Fe/H].

In \citet{Banks23} we performed a broader investigation into this spectro-seismic connection. We used moderate-resolution X-Shooter \citep{Vernet11} spectra of 49 red giant stars covering a tenfold larger wavelength range (i.e. $0.33-2.5\mu$) compared to \cite{Hawkins18}. We followed a similar strategy to that used by \cite{Hawkins18}, and used \cannon\ to generate a model trained on the spectra of the 49 stars. This model predicts the flux at each wavelength pixel as a quadratic function of stellar and asteroseismic labels, specifically \teff, \logg, [Fe/H], \dnu, and \texttt{RC\_Prob} (the seismic probability class prediction determined through the neural network classifier detailed in \citealt{Hon18}). From this investigation, similarly to \cite{Hawkins18}, we also found that molecular features, particularly CN and CO, are the most useful in classifying RC stars from RGB stars.

\begin{table*}
    \caption{\textit{Gaia} ID numbers, on-sky coordinates, observation date, exposure time, SNR, $W_2$ apparent magnitude, stellar parameters \teff, \logg, [Fe/H] from \citet{Buder21}, asteroseismic parameter \dnu\ from \citet{Reyes22}, and evolutionary state ($0=\rm{RGB}$; $1=\rm{RC}$) from \citet{Hon18,Hon22} for the stars in our data set. The full table is available in the online version of this article; an abbreviated version is included here to demonstrate its form and content.}
    \begin{tabular}{rrrlrrrrrrrr}
        \hline
          \textit{Gaia} ID & $\alpha(^{\circ})$ &  $\delta(^{\circ})$ &  UT Date &  $t_{\rm{exp}}$ (s) &  SNR & $W_2$ (mag) & \teff\ (K) &  \logg &   [Fe/H] &  \dnu\ ($\mu$Hz) &  RC/RGB \\
        \hline
        4760861403829025408 & 76.902 & -60.650 & 09-Nov-2020 &   2400 &  107 &   8.171 &  4575 & 2.453 & -0.492 & 4.59 &       0 \\
        4661624260335390464 & 73.677 & -67.584 & 24-Dec-2020 &   1800 &  111 &   8.227 &  4954 & 2.417 & -0.391 & 4.16 &       1 \\
        4759628267178723072 & 82.748 & -59.956 & 26-Dec-2020 &    600 &   99 &   8.095 &  4748 & 2.416 & -0.152 & 5.71 &       0 \\
        4653531614211520256 & 67.261 & -72.425 & 05-Nov-2020 &   1800 &  118 &   7.493 &  4923 & 2.400 & -0.405 & 4.23 &       1 \\
        4676446840986547328 & 64.583 & -63.034 & 02-Nov-2020 &   1800 &  127 &   7.979 &  4548 & 2.404 &  0.089 & 4.08 &       1 \\
        \hline
    \end{tabular}
    \label{tab:dataset}
\end{table*}

In addition to the CN, CO and CH features, we found that some atomic features also hold some significance, including iron, titanium and the iron peak elements vanadium and nickel. However, there are covariances in these features with other stellar labels, namely \logg\ and \teff\ in particular. Therefore, those atomic features may not be as important as \cannon\ model suggests in predicting the evolutionary state of red giants compared to other stellar parameters. 

In this study we investigate these features further, utilising a larger sample of red giant stars with known classifications from \tess\ asteroseismology ($N=123$) and high-resolution spectra obtained with the {\it Veloce Rosso} spectrograph at the Anglo-Australian Telescope (AAT) \citep[][]{Gilbert18}. Utilising a higher spectral resolution allows us to explore in more detail the spectral features identified in \cite{Banks23}, and potentially identify additional classification features not resolved in the moderate-resolution ($R\sim10,000$)  spectra of the previous study. We discuss this in Section \ref{subsec:finding_features}.

In addition, we analyse the spectral features that hold the most significance in predicting RC/RGB evolutionary state, thus providing insight into the astrophysical processes that drive the spectro-seismic connection of red giants. In particular, we focus on the photospheric abundances of carbon-bearing molecules and the $^{12}$C/$^{13}$C isotopic ratio, which are influenced by the first dredge-up, deep mixing along the RGB, and the helium flash. This is explored in Section \ref{sec:features}.

\section{Data} \label{sec:Data}

\begin{figure}
    \centering
    \includegraphics[width=0.481\textwidth]{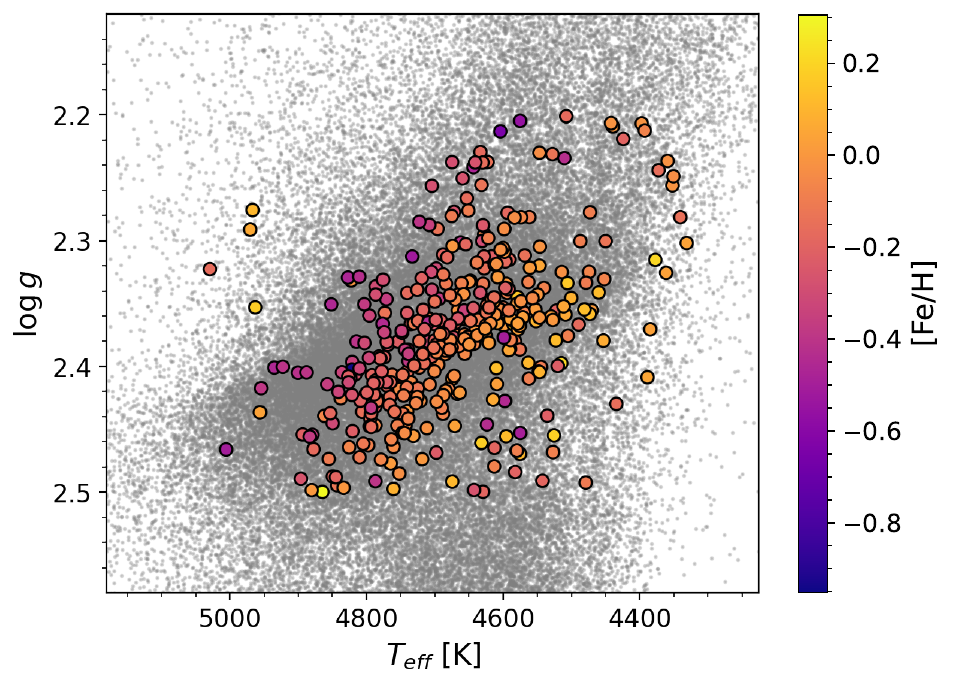}
    \caption{Kiel diagram of the stars in this data set overlaid on the GALAH DR3 sample \citep{Buder21}.}
    \label{fig:dataset}
\end{figure}

We made an initial collection of stars selected from a list of red giants 
collated from the \ktwo- and \tess-HERMES surveys \citep[][respectively]{Sharma19,Sharma18}
with effective temperatures $4300<T_\text{eff}<5100$\,K, surface gravity $2.2<\log g<2.5$, and metallicity $-0.7<\rm{[Fe/H]}<0.3$. 
These stars have also been explored with asteroseismology data from both the \ktwo\ mission \citep{Howell14} and the \tess\ mission \citep[][]{Ricker15} and their evolutionary states have been determined by \cite{Hon18,Hon22}. 
We made observations for a total of 301 stars (see Fig. \ref{fig:dataset}) with the \textit{Rosso} camera of the {\it Veloce} spectrograph \citep[$5800-9500$\,\AA, $R\sim80,000$;][]{Gilbert18} in conjunction with the 4\,m Anglo-Australian Telescope (AAT) at Siding Spring Observatory over the course of two observing semesters, 2020B and 2021B. Table \ref{tab:dataset} is a representative list of \textit{Gaia} IDs, on-sky coordinates, observation date, exposure time, $W_2$ apparent magnitude, stellar parameters (\teff, \logg, [Fe/H]) and asteroseismic labels (\dnu\ and evolutionary state, $0=\rm{RGB}$ and $1=\rm{RC}$) for the stars in our data set. A complete table detailing all red giants observed in our sample is included in the supplementary online material of this paper.

{\it Veloce} collects light from a 2.5\arcsec\ diameter aperture, using a 19-element hexagonal integral field unit (IFU), and uses optical fibres to reformat that in a 19-fibre linear pseudo-slit at the entrance of the echelle spectrograph. These 19 ``star'' fibres are supplemented by a ThXe simultaneous calibration exposure at one end of the pseudo-slit, a single-mode laser comb simultaneous calibration fibre at the other end of the slit, and 5 ``sky'' fibres offset significantly on-sky from the target star fibres. 

Each star was observed in a similar manner to observations collected in \cite{Palumbo22} in a series of multiple 600s exposures, such that the total exposure time (given the conditions) delivered an S/N per pixel of $>$50 at $\sim6670\,$\AA. For example, a star with an apparent $G$-band magnitude of 9.5 mag in typical AAT 1.5" seeing required 3$\times$600s exposures, while a $G=11$ star required 6$\times$600s exposures. At the middle of each 600s exposure, a 0.5s laser comb exposure was exposed as part of the standard {\it Veloce} calibration process. The laser comb inserts $\sim10,000$ diffraction-limited calibration lines into a single-mode fibre at the end of the {\it Veloce} pseudo-slit. Periodic ThXe exposures are obtained as well, and the observed `distance' on the detector between the laser comb and ThXe spots that result is used to calibrate the (slow) time-variation of the apparent length of the pseudo-slit.

The extraction of spectra from the {\it Veloce Rosso} echellogram uses a parametric spectrograph model, the parameters of which (the effective echelle grating spacing, the effective cross-disperser grating spacing, the pseudo-slit linear plate-scale, the detector X,Y zero-point position and detector rotation) are determined from each laser comb exposure, by doing astrometry using DAOPHOT on the individual laser comb lines and treating them as ``point sources''. These X,Y positions determine the spectrograph parameters for each exposure. Those parameters are then smoothed and interpolated as a function of time, to enable a spectrograph model to be derived for {\em any} intervening exposure (even if a laser comb was not obtained, as -- unfortunately -- does sometimes happen).

The spectrograph model for each observation then predicts the fibre tracks on the detector for each pseudo-slit fibre in each echellogram order, driving an optimal extraction for all the fibres in all the orders. The optimal extraction requires model fibre profiles for each fibre in each order as a function of Y (i.e. the high-resolution dispersion direction) position, and these are obtained by fits to flat-field observations in 60x100 pixel chunks. The environmental stabilisation of the spectrograph ensures that these profiles do not change with time. 
This final spectrograph model is then used to extract the spectra from the CCD images and write them into 3D data cubes (Y $\times$ fibre number $\times$ echelle order). The spectrograph model also provides a wavelength of each pixel in each order and in each fibre, and these are used to calibrate the wavelength of the optimally extracted spectra.

Because the spectrograph model is based on grating equations and a model for the relevant angles within the spectrograph, a first-order blaze function can be constructed based on the sinc$^2$ function for the relevant echelle grating diffraction angle. A polynomial correction to this ``ideal'' sinc$^2$ blaze function was constructed from hundreds of nights of flat-field data, to produce a single ``master blaze'' that can be applied to all data. 

Following this, the data is then scrunched to a common constant-velocity-spacing grid and pixels associated with telluric absorption (using the telluric map from NASA Planetary Spectrum Generator for a typical AAT exposure) deeper than 2 per cent are masked as bad.

These processing steps result in extracted exposure files with individual spectra from the 19 star fibres, five sky fibres, one ThXe calibration lamp fibre and one laser comb fibre, for each of the 40 spectral orders covering the wavelength range $5800-9500$\,\AA.

Following data extraction, each fibre is normalised by relative fibre throughput. Then an average sky spectrum is determined for each order by averaging the flux from the sky fibres and subtracted from the flux in each star fibre. The sky-subtracted star flux in each fibre is then normalised and combined into an average stellar spectrum for each order followed by merging each subsequent order. Merging the orders of echelle spectra often results in a distinct ripple shape in the merged spectra \citep[e.g.,][]{Cretignier20,Rozanski22}. Estimating a pseudo-continuum using a low-order polynomial to normalise the flux is non-trivial. Therefore, we made use of the neural network normalisation tool \textsc{SUPPNet} \citep{Rozanski22}. This tool filters through the domain of spectrum measurement to the domain of possible pseudo-continua using a fully convolutional neural network based on the semantic segmentation problem. Utilising this tool in our data reduction process successfully removed the distinct ripple shape and resulted in normalised continua $\sim1$.

The final step of data reduction, before the spectra are in a suitable state for use and analysis in \cannon, is performing velocity corrections. First is a heliocentric correction followed by correcting for the stars' radial velocity via a Doppler correction. We achieve this with cross-correlation via the \texttt{crosscorrRV} function from the \textsc{PyAstronomy} python package \citep{pyasl}. We choose one star in the data set with an estimated relative velocity close to zero from the GALAH DR3 catalogue (\texttt{star\_id}: 05185082-5707494, $\rm{RV}=0.0187\,\rm{km/s}$) to be the template spectrum. The remaining spectra, i.e. the observation spectra, are masked to a spectral window between $8480$\,\AA\ and $8680$\,\AA\ where there are numerous strong features present in all spectra such as the Ca II infrared triplet. The template spectrum is then Doppler shifted across a range of possible relative velocities ($-150<\rm{RV}<250\,\rm{km/s}$ in steps of $0.1\,\rm{km/s}$) and linearly interpolated to the wavelength points of each observation spectrum to calculate the cross-correlation function:

\begin{align}
    CC(v_j)=\sum^{N}_{i=1}\alpha_i\times(f_i \times t(w_i -\Delta_{i,j}))
\end{align}
for $N$ data points $f_i$ at wavelengths $w_i$ depending on the velocity $v_j$ and weights $\alpha_i$, where $\Delta_{i,j}=w_{i}(v_{j}/c)$. The weights used in this calculation are simply the inverse square of the flux uncertainties.

The relative velocity between the template spectrum and an observation spectrum is the relative velocity where the cross-correlation is at a global maximum. We then Doppler shift the unmasked observation spectra with respect to the determined relative velocity, which results in different wavelength arrays for each spectrum. \cannon\ requires all spectra to be on the same wavelength grid, so we perform a linear interpolation of all spectra onto a common wavelength grid.
Following Doppler correction, we mask common regions across all spectra that exhibit telluric features.

Following these steps, the spectra are then in a suitable state for use and analysis in \cannon\ as well as for direct comparison of spectral features for the investigation explored in Section \ref{sec:features}.

In addition to normalised spectra on a common wavelength grid, \cannon\ also requires a precise estimate of the flux variance for each pixel of the spectra. 
We assume the flux uncertainty to be Gaussian and hence use it to weight the influence of certain pixels when computing the best set of labels via minimum $\chi^{2}$ estimation. For computational speed, \cannon\ is fed with the inverse variance or inverse of the squared flux uncertainty.
Another necessary input to establish a spectral model with \cannon\ are labels that describe the spectra to train the model, for example, stellar parameters such as \teff, \logg, and [Fe/H]. In this investigation, we utilise those stellar parameters from GALAH DR3 \citep{Buder21}. GALAH \teff\ is estimated from the spectra rather than via photometry with precision to 49\,K and \logg\ is estimated via bolometric relations. \cite{Buder21} find excellent agreement with the values of \logg\ for Gaia Benchmark Stars as well as those that are cross-validated with asteroseismology. Finally, the [Fe/H] abundances reported in GALAH DR3 are determined from a global abundance of [Fe/H] from all Fe lines in the GALAH DR3 line list, resulting in a precision of 0.055 dex.

\begin{figure}
    \centering
    \includegraphics[width=0.481\textwidth]{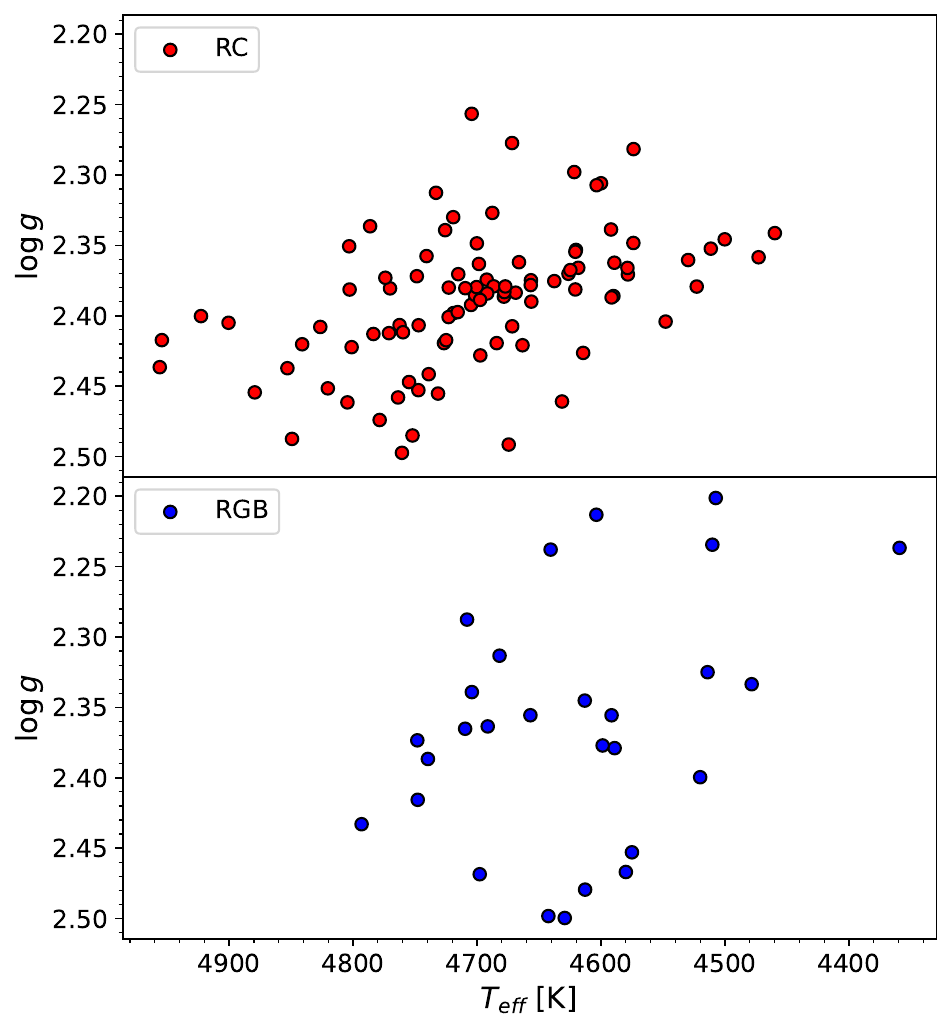}
    \caption{Kiel diagram of the stars identified as RC in the top panel and RGB in the bottom panel. The stellar parameters \teff\ and \logg\ are sourced from GALAH DR3 \citep{Buder21} and the RC/RGB classifications are determined using the neural network classifier detailed in \protect\cite{Hon17}.}
    \label{fig:RC-RGB-keil}
\end{figure}

We also incorporate asteroseismic information derived from \tess\ asteroseismology. A total of 123 stars in our data set (shown in Figure \ref{fig:RC-RGB-keil}) have reliable asteroseismic parameters available for our analysis. These include the large frequency separation, \dnu, and their evolutionary state classification. We use the \dnu\ values presented in \cite{Reyes22}. These values are determined for red giants with at least six months of \tess\ observations via the SYD asteroseismic pipeline \citep{Huber09,Huber11,Yu18}. \cite{Reyes22} vet those values and determine a reliability score using a neural network classifier. This ensures that \dnu\ estimates are reliable regardless of the method used for their estimation. The evolutionary states of these red giants have been identified in \cite{Hon18} and \cite{Hon22}.

\section{Analysis with \cannon} \label{sec:Cannon}
\cannon\footnote{\cannon\ is named after Havard Computer Annie Jump Cannon as a recognition for her pivotal work in classifying stars without the need for stellar spectral models.} \citep{Ness15,Ho16,Casey16} is a powerful tool that allows for the prediction of stellar parameters from observed stellar spectra. It is a data-driven algorithm that predicts stellar parameters without relying on a grid of synthetic spectra, unlike other tools that employ a physics-based approach by fitting an observed spectrum to a synthetic spectrum. One particular advantage of \cannon\ is its ability to predict stellar labels from lower signal-to-noise data with comparable accuracy to current physics-based approaches.

\cannon\ works in essentially two steps: the \textit{training step} and the \textit{test step}. The \textit{training step} involves creating a generative spectral model from a set of input spectra and the stellar labels that describe those spectra (i.e. the \textit{training objects} and \textit{training labels} respectively). This model describes a probability density function (pdf) for the flux at every pixel in the spectrum as a function of input labels. Following the \textit{training step}, the \textit{test step} assumes this generative model holds for all other objects in the data set (i.e. the \textit{test objects}). The spectra of the \textit{test objects} are each compared to the generative model, which allows for the labels of the \textit{test objects} to be solved providing, in effect, a label transfer from the \textit{training objects} to the \textit{test objects}. See \cite{Ness15} for a more in-depth explanation. 

\begin{figure}
    \centering
    \includegraphics[width=0.481\textwidth]{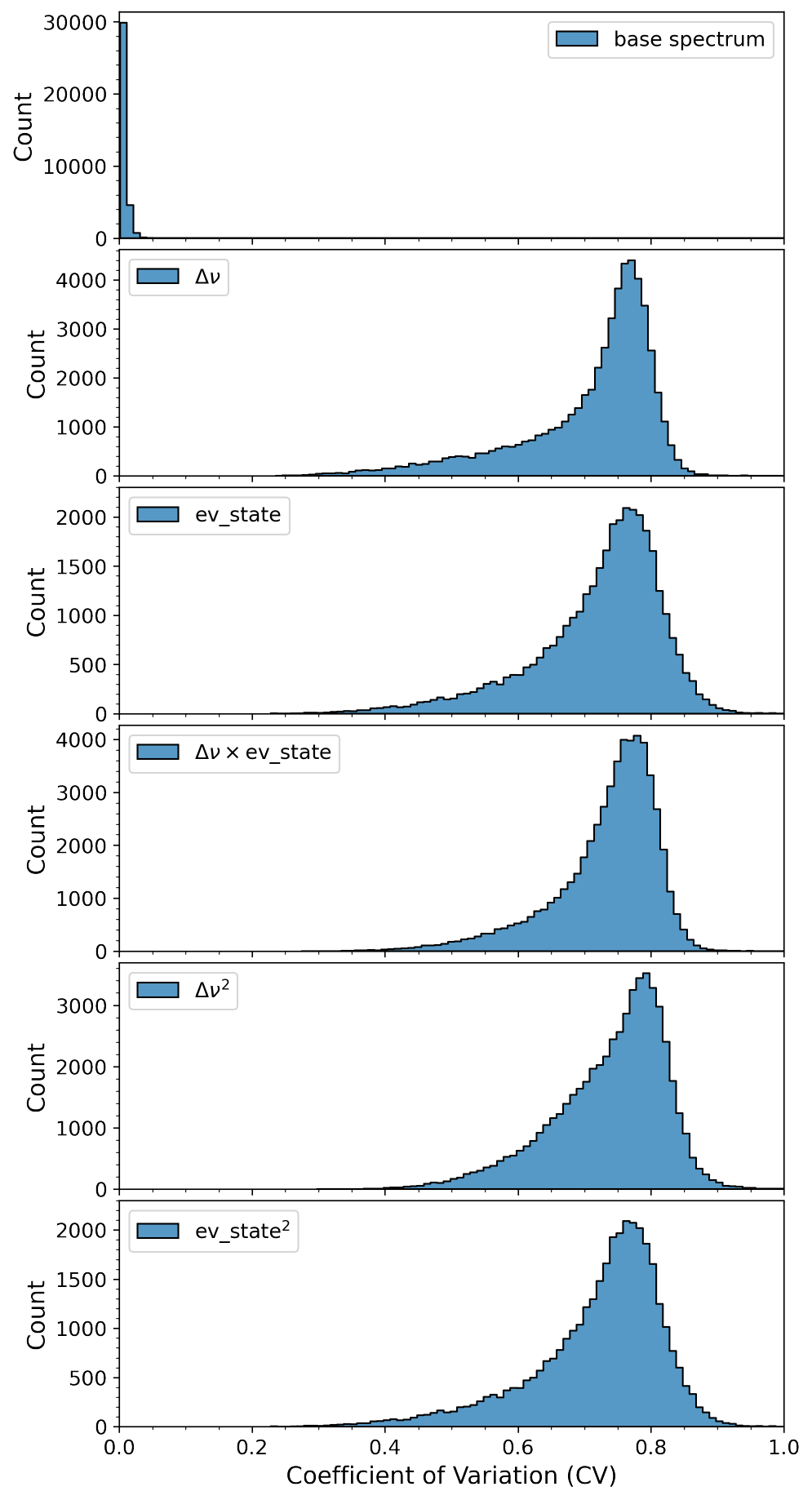}
    \caption{The distribution of the coefficient of variation (CV) for each model coefficient across all models. The top panel represents the base spectrum of the models and the following panels represent \dnu, ev\_state, $\Delta\nu\times\rm{ev\_state}$, \dnu$^{2}$, and ev\_state$^{2}$ terms of the bootstrap models respectively. The CV for each model coefficient peaks below 1, denoting a low variation across the different bootstrap models.}
    \label{fig:theta_CV}
\end{figure}

\subsection{Validating \cannon\ Model} \label{subsec:validation}
First, we conduct a bootstrap analysis to assess the reliability of \cannon\ and to ensure consistent results in identifying significant spectral features for classifying red giants, irrespective of the selection of stars used to build the spectral model.
We generated $1,000$ training sets made of the spectra of random selections of 18 RGB stars and 54 RC stars in the data set with known asteroseismic classifications. This results in a total training set size of 72 i.e. $\sim60$ per cent of the total number of red giants with known asteroseismic classifications and is representative of the ratio of RGB and RC stars in that sample. These training sets were used to train $1,000$ \cannon\ models with the asteroseismic labels \dnu\ and evolutionary state, as well as the stellar parameters \teff, \logg, and [Fe/H], to predict the flux at each pixel as a quadratic function of these labels.

We investigate the variability of the individual model coefficients ($\boldsymbol{\theta_{n}}$), with a particular focus on the asteroseismic labels. This is achieved by calculating the coefficient of variation \citep[CV;][]{Koopmans64} of each pixel for each $\boldsymbol{\theta_{n}}$ across all models. The CV measures the variability of each model coefficient at each pixel across all models, normalised by the mean, providing insight into the impact of the selection of the training set.
We choose to focus on the asteroseismic labels in our validation because it has previously been shown that \cannon\ can predict stellar labels (i.e., \teff, \logg, and [Fe/H]) with high accuracy \citep[e.g.,][]{Ness15,Hawkins17}{}{}. 
The top panel of Figure \ref{fig:theta_CV} represents the CV for $\boldsymbol{\theta_{0}}$ which is the ``base spectrum'' of \cannon\ model \citep{Ness15}. The following panels represent the CV for the remaining model coefficients pertaining to the asteroseismic labels: \dnu, ev\_state, $\Delta\nu\times\rm{ev\_state}$, $\Delta\nu^{2}$, and ev\_state$^{2}$ respectively. 
The distribution of the CV for each pixel across all bootstrap models peaks below 1, indicating low overall variation (i.e. the standard deviation of the distribution of the model coefficients is less than the mean).

Therefore, we can confidently expect consistent results in identifying significant spectral features for the classification of red giants, irrespective of the selection of RC and RGB stars for the training set. In addition, we are confident that the coefficients of the model we use in Section \ref{subsec:finding_features} are meaningful, and that the spectral features we identify as significant based on the coefficients do truly correspond to whether a star belongs to the RC or RGB. In Appending \ref{sec:predicting_state} we provide more detail on the RC/RGB classifications produced by \cannon\ and how those are affected by the selection of the training set. We further compare the reliability of these classifications to alternative methods in the literature such as those based on photometry and stellar parameters.

\begin{figure*}
    \centering
    \includegraphics[width=\textwidth]{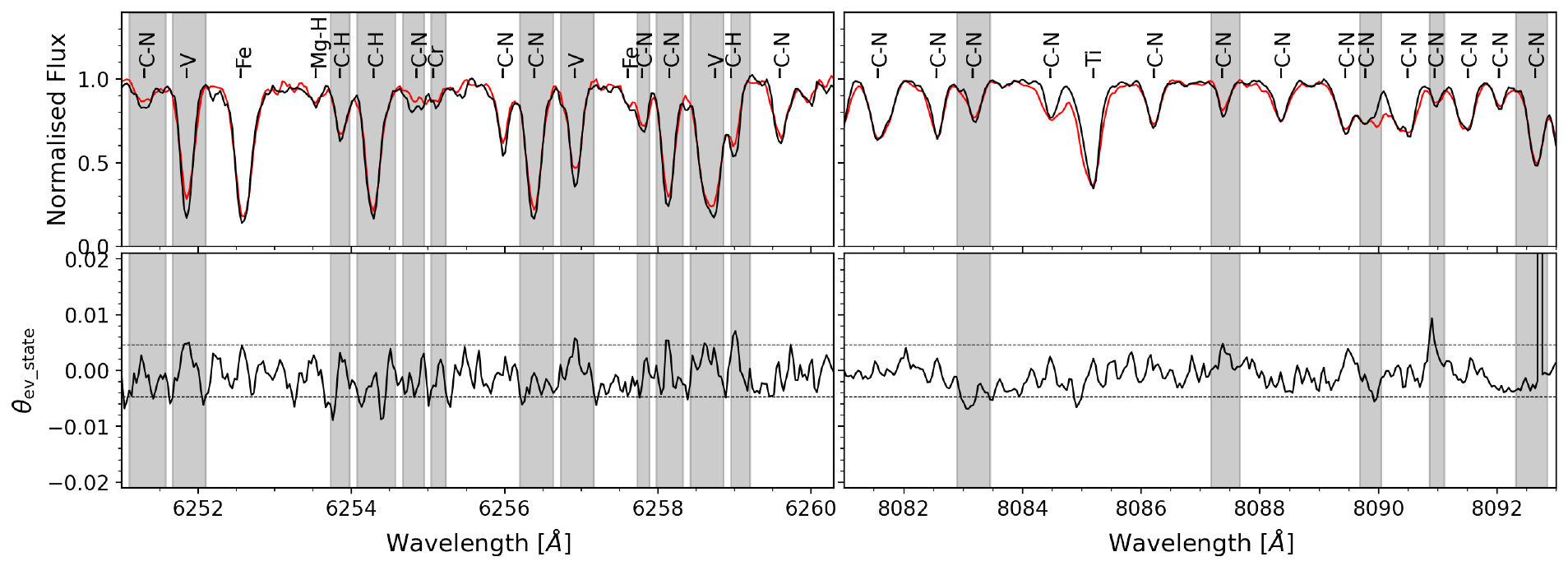}
    \caption{A representative selection of spectral features \cannon\ identifies as the most significant in determining the evolutionary state of red giant stars in two spectral windows (i.e. $\sim6256\pm5$\,\AA\ and $\sim8087\pm6$\,\AA). The upper panels show the flux of red giants in our data set with labelled spectral features. The red line represents the flux of one RC star in the data set (\textit{Gaia} DR3 ID: 4661172395405681920, $T_{\rm{eff}}=4511$, $\log g=2.35$, and $\rm{[Fe/H]}=0.05$). The black line represents the flux of one RGB star in the data set (\textit{Gaia} DR3 ID: 4766063055901392128, $T_{\rm{eff}}=4514$, $\log g=2.32$, and $\rm{[Fe/H]}=-0.02$). The solid black line in the lower panels represents the model coefficients of the evolutionary state label (ev\_state) for each pixel. The dotted lines about $\pm0.0047$ represent the 90th percentile of the distribution of the ev\_state model coefficient. We have highlighted in grey the spectral features in these spectral windows that we initially identified as significant (without regard to any covariances with other labels).}
    \label{fig:initial_significant_features}
\end{figure*}

\subsection{Identifying Significant Spectral Features} \label{subsec:finding_features}

We initially identify prominent spectral features in the spectra of our data set using the find\_lines\_derivative function from the \textsc{specutils} python package \citep{specutils19}. This method identifies emission and absorption features in a spectrum based on finding zero crossings in its derivative, thus indicating the bottom of an absorption feature or the top of an emission feature.

Once spectral features are identified within the spectra, we then cross-reference those wavelengths with a line list for the wavelength range of the \textit{Rosso} camera of the \textit{Veloce} spectrograph ($5800-9500$\,\AA), created with the atomic and molecular line list generator \textsc{linemake} \citep{Placco21}. In the high-resolution spectra of the red giants in our data set, we identify a total of 1,823 spectral absorption features. The majority of these features are CN (56 per cent), Fe (12 per cent), CH (4 per cent) and Ti (4 per cent). 

To determine which spectral features are significant in the prediction of red giant evolutionary state, we train \cannon\ with the spectra of all stars with asteroseismic information (95 RC; 28 RGB). We establish a spectral model with \cannon\ that predicts the flux of red giants as a quadratic function of \teff, \logg, [Fe/H], \dnu\ and RC/RGB evolutionary state. This allows us to investigate the values of the coefficients for the linear and quadratic terms pertaining to the evolutionary state of red giants and determine which pixels are most useful in their classification. 

\begin{figure}
    \centering
    \includegraphics[width=0.481\textwidth]{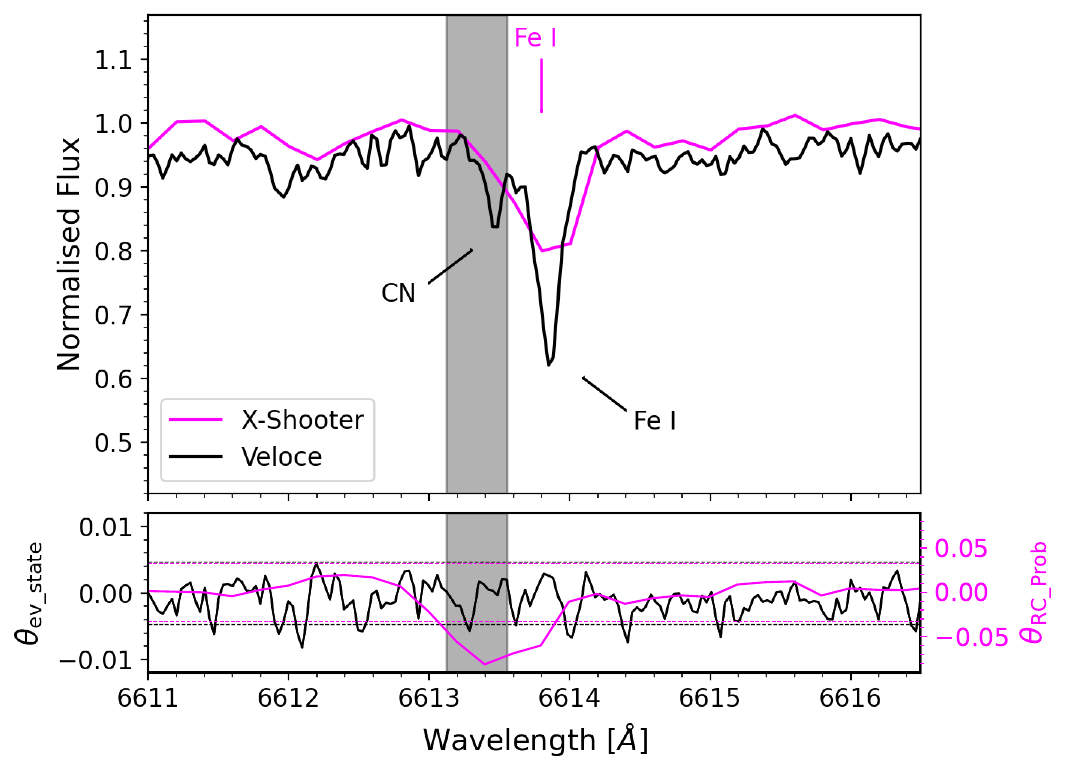}
    \caption{The top panel shows a comparison between a moderate-resolution X-Shooter spectrum (magenta) and high-resolution \textit{Veloce Rosso} spectrum (black) for two RC stars with similar stellar parameters (\teff$\sim4715\,$K, \logg$\sim2.4$ and [Fe/H]$\sim-0.2$). In the moderate-resolution X-Shooter spectrum, only the Fe I line is able to be resolved; however, in the high-resolution \textit{Veloce Rosso} spectrum, this blended line is separated into the Fe I line and a CN molecular line. In the bottom panel, we show the model coefficients for the linear terms pertaining to the \texttt{ev\_state} label (black) and the \texttt{RC\_Prob} (magenta) label for \cannon\ models trained on \textit{Veloce Rosso} and VLT/X-Shooter spectra respectively. Here we also show the 90th percentile of the distribution of each model label (i.e. the threshold for significance used in both this study and \citealt{Banks23}) in each respective colour as dotted lines. In this analysis, it is clear that the significant feature pertaining to the prediction of red giant evolutionary phase is the highlighted CN line and not the Fe I line.}
    \label{fig:xshooterVSveloce}
\end{figure}

We initially define a spectral feature as significant to the prediction of red giant evolutionary state following a similar method to that in \citet{Banks23}. 
A spectral feature is significant if any of the pixels that are part of the line and not the surrounding continuum exhibit an ev\_state model coefficient greater than the 90th percentile of the distribution ($\pm0.0047$).
This is effective as a first identification since the linear term model coefficients represent the lowest order dependence of the flux on the respective label \citep{Buder21}. We further refine our selection to require that significant spectral features must also exhibit an ev\_state$^{2}$ model coefficient value greater than the 90th percentile of the distribution of the ev\_state$^{2}$ coefficients ($\pm0.03$). This results in a total of 504 identified significant features. 

Figure \ref{fig:initial_significant_features} illustrates a representative selection of significant features that hold the most information regarding the evolutionary state of red giants according to their ev\_state and ev\_state$^{2}$ coefficients across two spectral windows ($\sim6256\pm5$\,\AA\ and $\sim8087\pm6$\,\AA). In the upper panels, we show spectra for one RC star in red (\textit{Gaia} DR3 ID: 4661172395405681920, \teff$=4511$, \logg$=2.35$, [Fe/H]$=0.05$) and one RGB star in black (\textit{Gaia} DR3 ID: 4766063055901392128, \teff$=4514$, \logg$=2.32$, [Fe/H]$=-0.02$). The solid line in the lower panels shows the linear ev\_state model coefficient value for each wavelength pixel, while the dashed lines at $\pm0.0047$ illustrate the 90th percentile of the distribution of the ev\_state linear model coefficient. Spectral features that are identified as significant to the prediction of red giant evolutionary state are highlighted in grey. 

The majority of spectral features that we identify as significant are CN molecular features, followed by Fe and CH. A total of 42 spectral features overlap with those found in the moderate-resolution VLT/X-Shooter spectra from our previous broad wavelength investigation \citep{Banks23}. On visual inspection of the \textit{Veloce Rosso} spectra we can see that 18 of those 42 overlapping spectral features are actually blended in the VLT/X-Shooter spectra. With the benefit of higher resolution, we can see that the significant feature is the feature we identified in \citet{Banks23} in 5 of 18 cases, but in the other 13 cases, we can see that the significant feature is due to a different transition (one example is shown in Figure \ref{fig:xshooterVSveloce}). Lines with new, corrected identifications in our final line list of significant spectral features are noted in Appendix \ref{appendix:B}.

\begin{figure}
    \centering
    \includegraphics[width=0.481\textwidth]{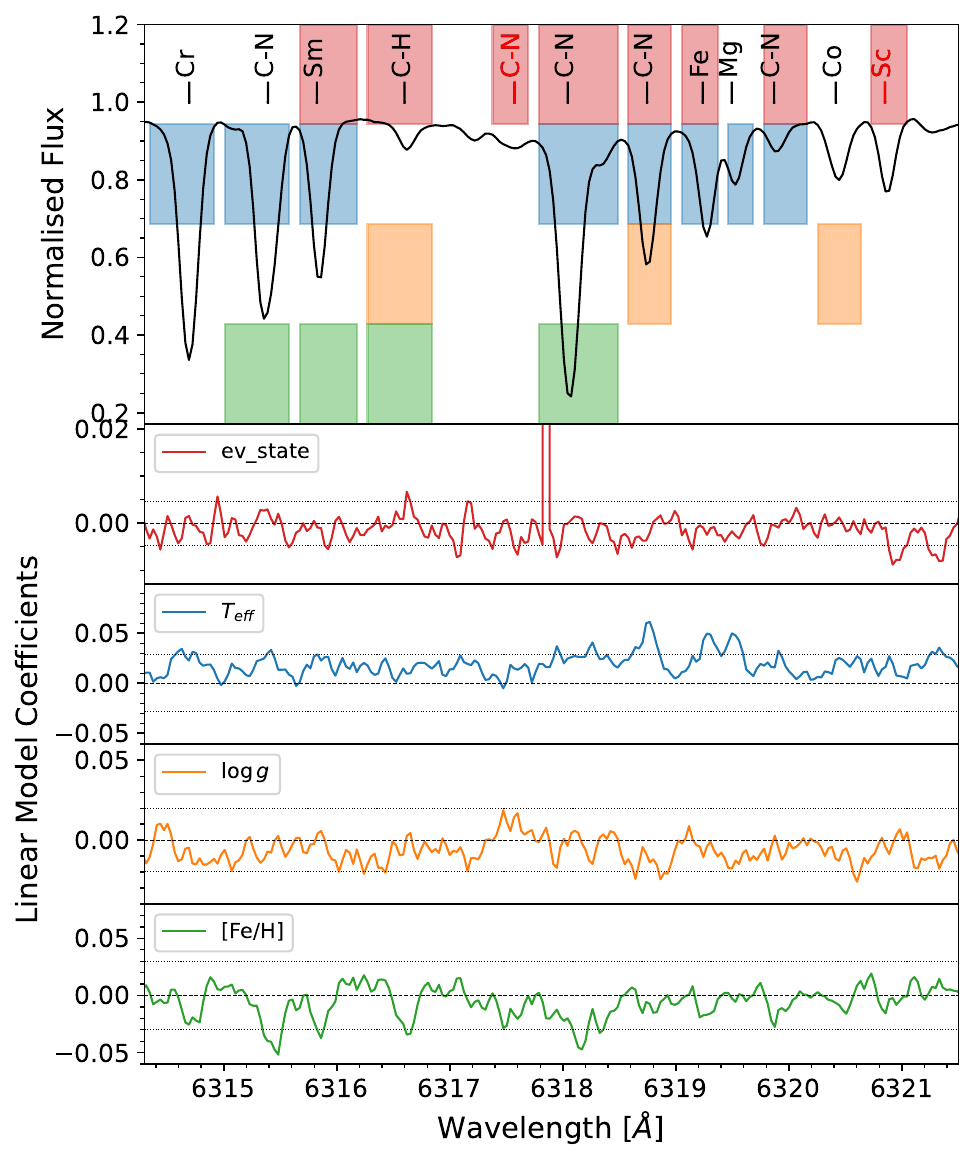}
    \caption{A representative selection of absorption features in the spectral window near $\sim6318\pm4$\,\AA\ and the corresponding model coefficients for the linear terms ev\_state, \teff, \logg, and [Fe/H] in the lower panels (red, blue, orange, and green lines respectively). The top panel illustrates the base spectrum of \cannon\ model with the major features labelled. We demonstrate whether a spectral feature is defined as significant to predicting a particular label with coloured bars. Those that are highlighted in red are significant in predicting red giant evolutionary state, blue for \teff, orange for \logg, and green for [Fe/H]. In our final selection of significant spectral features that hold information about the evolutionary state of red giants, we select those that are only significant in the prediction of the evolutionary state and not any other stellar label, such as the CN lines at $\sim6318$\,\AA\ (whose labels are highlighted in red).}
    \label{fig:covariances}
\end{figure}

We demonstrate this difference in Figure \ref{fig:xshooterVSveloce}. Here we show a spectral window centred on the Fe I line at $\sim6614$\,\AA\ for an RC star with a moderate-resolution X-Shooter spectrum in magenta and a different RC star with a high-resolution \textit{Veloce Rosso} spectrum in black. These stars have similar stellar parameters: \teff$\sim4715\,$K, \logg$\sim2.4$, [Fe/H]$\sim-0.2$. It is evident from the \textit{Veloce Rosso} spectrum that the Fe I line in the X-Shooter spectrum is blended with the highlighted CN line. The higher-resolution \textit{Veloce Rosso} spectra enable us to distinguish the previously blended lines and identify that the significance is actually in the CN feature.

In our previous investigation \citep[][]{Banks23}{}{}, we did not probe potential covariances these significant features may have with other stellar labels. For example, a spectral feature identified as significant in the prediction of red giant evolutionary state may also be significant in the prediction of other stellar labels. The reason to include stellar parameters in \cannon\ model is to detect and factor out their effects on the spectra of red giants so that the influence of red giant evolutionary state can be extracted as cleanly as possible. However, the quadratic model in \cannon\ is not a perfect representation of stellar spectra, and we cannot completely separate the effects of abundance from the effects of \teff, \logg, and [Fe/H] on line strengths. 

To be as certain as possible that we are focusing on features that are significant for the prediction of red giant evolutionary state, we further restrict our list to only consider the lines that are significant for \texttt{ev\_state} but are not significant for the stellar parameters. 
Of the initial 504 spectral features we found to be significant in the prediction of red giant evolutionary state, 293 ($\sim58$ per cent) are also significant in predicting \teff, 197 ($\sim39$ per cent) are also significant in predicting \logg, and 244 ($\sim48$ per cent) for the prediction of [Fe/H].

The top panel of Figure \ref{fig:covariances} shows the base spectrum of \cannon\ model, i.e. a representative spectrum for a star with $T_\text{eff}=4330\,$K, $\log g=2.2$, $\text{[Fe/H]}=-0.95$, $\Delta\nu=2.80$ and $\texttt{ev\_state}=0$, in the spectral window $\sim6318\pm4$\,\AA\ with prominent spectral features labelled. The lower panels show the model coefficients for the linear terms for the labels \teff\ in blue, \logg\ in orange, [Fe/H] in green and evolutionary state in red. Spectral features that are significant to the prediction of each of these labels are highlighted in their respective colour in the upper panel. 
For example, the CH line at $\sim6316.7$\,\AA\ holds high significance in the prediction of red giant evolutionary state as well as \logg\ and [Fe/H].
Requiring that significant lines are only significant for \texttt{ev\_state} reduces our list to a total of 66 spectral features, which are again dominated by CN, including the CN lines at $\sim6317.5$\,\AA. For a full line list see Table \ref{tab:linelist}.

\section{Spectral Feature Analysis} \label{sec:features}
Abundance determination for the stars in this data set is the topic of upcoming work, but there are useful indicators of abundance, isotopic ratio, and the physical mechanisms responsible for the spectro-seismic connection in the equivalent widths of significant features and \cannon\ coefficients we are considering in this study. Our RGB stars were selected to have similar stellar parameters to our RC stars, which are fairly confined by nature. As a result, the relative strengths of absorption lines (as measured by their equivalent widths), which are the important factor for \cannon, reflect abundances more directly than they would for a more heterogeneous set of stars, and we can use them as a rough proxy to compare the abundances between the RC and RGB stars. However, our data set does cover a range of 1 dex in [Fe/H], 800K in \teff\ and 0.3 in \logg, and this will add some scatter to the mapping from abundance to line strength. We expect that any trends found in the spectral feature analysis will become clearer when viewed as abundance trends. 

Within the 66 spectral features, we find to be significant for predicting evolutionary state, we find that the linear coefficient for the ev\_state label in \cannon\ is negative for the majority of CN lines and positive for the majority of CH lines. This indicates that the CN lines are stronger and the CH lines are weaker in RC stars than in RGB stars, implying a higher nitrogen abundance and lower carbon abundance in the more evolved stars. This is the direction we expect for those abundances to evolve as a result of deep mixing during the RGB phase \citep[e.g., ][]{Gratton00, Martell08}.

The \Cratio\ ratio for carbon isotopes is also known to change as a result of deep mixing during the RGB phase \citep[e.g., ][]{Charbonnel98}. Its typical value drops from around 20 to around 8, and remains low in RC and horizontal branch stars.
To test whether this effect can be seen in our data set and whether the \Cratio\ ratio is meaningful for the spectro-seismic connection, we compared the ratio of line strength in \Ctwo\ and \Cthree\ features in RC versus RGB stars. As a result of \Ctwo\ depletion and \Cthree\ enrichment during deep mixing, a lower \Cratio\ ratio in RC stars should, therefore, result in a lower line strength ratio in those stars (modulo the effects of varying temperatures and metallicity within our data set). We identified three pairs of \Ctwo\ and \Cthree\ features within our spectra (that is, three transitions with a clear isotope shift) where at least one of the pair has been identified as significant in our \cannon\ model for the prediction of red giant evolutionary state (see Table \ref{tab:Cratio}). We use the \texttt{REvIEW} code \citep[Routine for Evaluating and Inspecting Equivalent Widths;\footnote{https://github.com/madeleine-mckenzie/REvIEW/tree/main}][]{McKenzie22}, an automated tool that fits individual absorption lines using the \texttt{scipy.optimize} function \texttt{curve\_fit}, to determine equivalent widths for those lines in the 123 stars for which we have asteroseismic classifications. 

\begin{figure}
    \centering
    \includegraphics[width=0.481\textwidth]{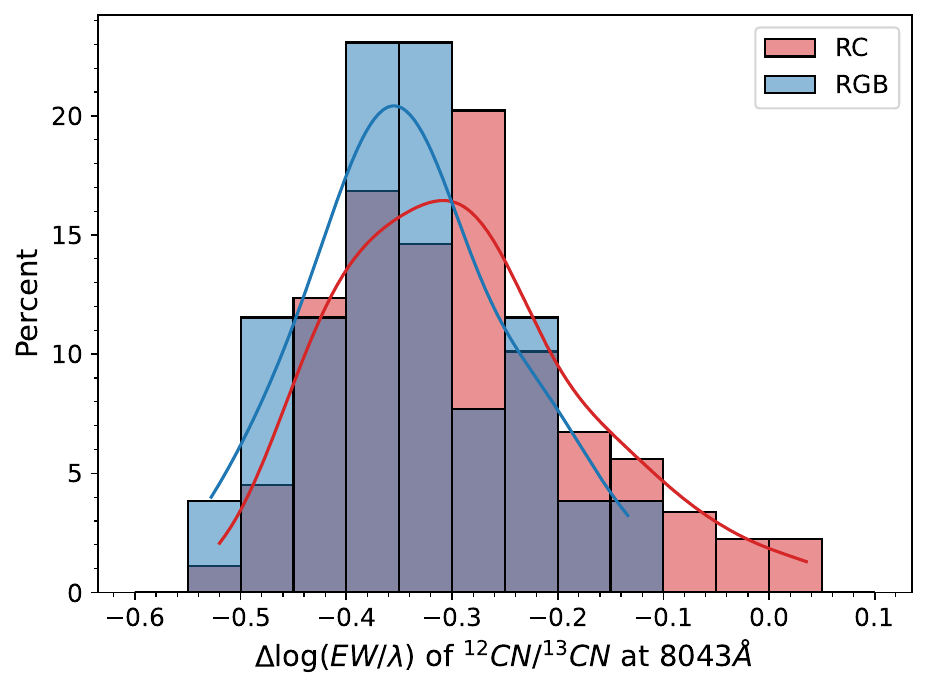}
    \caption{Histogram of the difference in reduced equivalent width between the \Ctwo\ and \Cthree\ lines near $8043\,$\AA, corresponding to the ratio of the two line strengths. RC stars are shown in red, and RGB stars are shown in blue. The red and blue solid lines are kernel density estimates for the same data. While both distributions are broad, there is a slight preference for RC stars to have a less negative value of $\Delta \log\frac{\textrm{EW}}{\lambda}$, indicating a difference of the \Cratio\ ratio in RC and RGB stars. This supports the notion that the surface abundance changes caused by deep mixing on the RGB are the main drivers of the spectro-seismic connection of red giants.}
    \label{fig:C12C13-ratio}
\end{figure}

\begin{table}
    \centering
    \caption{The locations of the three \Ctwo\ and \Cthree\ pairs within our spectra that we used to investigate inferences of the \Cratio\ ratio between RC and RGB stars in our sample. The isotopic transition from each pair that is identified as significant in our analysis is marked with an asterisk.}
    \label{tab:Cratio}
    \begin{tabular}{cc}
        \hline
        $\lambda$\,(\Ctwo) & $\lambda$\,(\Cthree) \\
        \AA\ & \AA\ \\
        \hline
        6317.568* & 6318.042 \\
        6668.378* & 6667.744 \\
        8043.311* & 8043.874 \\
        \hline
    \end{tabular}
\end{table}

To test for a difference in the \Cratio\ ratio between RC and RGB stars in our data set, we consider $\Delta \log \frac{\textrm{EW}}{\lambda}$ (\Ctwo$-$\Cthree), the difference between the reduced equivalent width for these isotopic line pairs, in the sense of \Ctwo\ minus \Cthree. Due to the depletion of \Ctwo\ and the enrichment of \Cthree\ along the RGB, we expect the population of RC stars to be slightly less negative than the RGB population. Figure \ref{fig:C12C13-ratio} shows this difference for the \Cratio\ line pair near 8043\AA\, across all 123 stars with asteroseismic classifications. There is a slight deviation in the distributions of the RC and RGB samples, suggesting a potential difference in the \Cratio\ between the two populations due to carbon depletion from deep mixing along the RGB.
We performed a two-sample Kolmogorov–Smirnov (KS) test to investigate whether the distributions of the difference in reduced EW for each line for RC and RGB stars are consistent with being drawn from the same parent distribution. This test returned a $p$-value of $2.3\times10^{-25}$, indicating that the two distributions are more likely to have been drawn from different source populations. 
However, additional abundance analysis is necessary to confirm whether the \Cratio\ follows the anticipated trends resulting from carbon depletion along the RGB.

Star-to-star differences in stellar parameters will add scatter to the line strengths even at fixed abundance, broadening both the RC and RGB distributions of reduced equivalent width difference. While this look at reduced equivalent widths is suggestive of a difference in the \Cratio\ ratio between our RC and RGB stars, isotopic abundances determined through spectrum synthesis are needed to be sure.

\section{Summary} \label{sec:Conc}

This investigation is the first high-resolution study of the spectro-seismic connection of red giant stars. Here we have used \textit{Veloce Rosso} spectra of 123 red giants to train a generative model with the data-driven algorithm \cannon. We trained a model with 95 RC and 28 RGB stars with known asteroseismic classifications from \tess\ asteroseismology \citep{Hon18,Hon22}. With this \cannon\ model, we were able to investigate which spectral features hold the most significance in predicting the evolutionary state of red giant stars, as well as probe these spectral features further to reveal more detail about the astrophysical processes that drive this spectro-seismic connection.

We followed a similar method for identifying significant spectral features for the prediction of red giant evolutionary state to that introduced in \cite{Banks23}. A spectral feature was determined to be significant in the prediction of red giant evolutionary state if the model coefficient pertaining to both the linear and quadratic terms for the evolutionary state label were greater than the 90th percentile of the distribution of those coefficients across all pixels. Following this same method as \cite{Banks23} we found 504 such features that are significant in the prediction of red giant evolutionary state. 

However, as noted in \cite{Banks23}, many of these significant features also hold some significance in the prediction of the other stellar labels (\teff, \logg, and [Fe/H). To be as certain as possible that we are focusing on features that are truly influenced by the evolutionary state of red giants, we refine our list of 504 significant features to exclude those that are also found to be significant in the prediction of other stellar labels. This reduces our list of significant features to 66, which are dominated by CN molecular features ($\sim53$ per cent). This is in agreement with previous studies \citep[e.g.,][]{Hawkins18,Banks23} that also find CN molecular features to be the dominant tracer of the spectro-seismic connection of red giant stars. 
In this process, we found that 42 of the significant features were also significant in the X-Shooter data we used for analysis in \citet{Banks23}. With the higher spectral resolution of \textit{Veloce Rosso} we found that 18 of those were actually blended lines, and for 13 of the 18 the feature with the most significance for evolutionary state was not the one we originally identified. In Appendix \ref{appendix:B} we have compiled a list of the 66 significant features that are only significant for predicting the evolutionary state of red giants, and we have noted the features for which our identification has changed since the previous study.

Focusing on the features we find to be significant in the prediction of red giant evolutionary state, we found that a stronger absorption in CN features and a weaker absorption in CH features was associated with \cannon\ coefficient for ev\_state for the majority of significant CN and CH features, suggesting that the typically higher [N/Fe] and lower [C/Fe] abundance in RC stars relative to RGB stars is an important contributor to the spectro-seismic connection. 

We also considered the relative strengths of absorption in CN isotopic pairs, i.e., the same transitions for \Ctwo\ and \Cthree. There are three such pairs of lines in our red giant spectra where at least one member of the pair is identified as significant only to the prediction of red giant evolutionary state. We measured the 
reduced equivalent widths for these lines across all our red giants with asteroseismic classifications using the automated EW estimator \textsc{review}. 
We then performed a two-sided KS test to investigate whether 
the distributions of the difference in reduced EWs between \Ctwo\ and \Cthree\ were significantly different between 
the RC and RGB populations. For the line pair near 8043\AA, this test returned a $p$-value of $2.3\times10^{-25}$, indicating that we cannot reject the hypothesis that the two distributions originated from two different parent populations.

These tendencies in line strength behave as we would expect if the physical processes that change the surface abundances and isotopic ratios in red giant stars, such as deep mixing and the helium flash, are the driving forces of the spectro-seismic connection in red giants. Follow-up work including elemental and isotopic abundance determination will help to solidify this connection and quantify whether red giants with similar \teff\ and \logg\ can be distinguished by quantifying the C and N abundances or measuring \Cratio\ without data-driven analyses such as those performed with \cannon.

\section{Discussion and Future Work} 

This work builds on previous investigations, first by \cite{Hawkins18} and \citet{Ting18} using APOGEE ($1.5-1.7\,\mu$, $\rm{R}\sim 28~000$) and LAMOST ($3600-9000\,$\AA, $\rm{R}\sim1800$) spectra for large samples of red giants, followed by a broad wavelength investigation by \cite{Banks23} with red giant spectra from the VLT/X-Shooter spectrograph ($0.33-2.5\,\mu$, R$\sim 10~000$). Molecular features involving carbon (i.e. CN, CO and CH) were found to be significant in the prediction of red giant evolutionary state in these studies; however, further investigation was required to understand which astrophysical processes could be responsible for the spectro-seismic connection. 

Using the coefficients of \cannon\ model for our 123 red giant stars with asteroseismic classifications, 
we find that the spectral features in our \textit{Veloce Rosso} sample that are significant for separating RC and RGB stars are likely to correspond to the surface abundance changes caused by deep mixing on the RGB. Specifically, the depths of CN and CH absorption lines, along with the line depth ratios for the same transitions in \Ctwo\ versus \Cthree\ correlate to \cannon\ coefficient for ev\_state. This suggests that RC stars are expected to display lower [C/Fe], higher [N/Fe], and a lower \Cratio\ ratio compared to RGB stars.

The accumulated result of ongoing abundance changes caused by the first dredge-up and deep mixing on the RGB must be visible in the abundances of RC stars, which have already experienced their full RGB lifetime plus the helium flash at the tip of the RGB. However, a detailed abundance analysis based on these same spectra is necessary to be sure that the differences we find in absorption line strength between RC and RGB stars is truly representative of the expected changes in abundance during red giant evolution. A study on this topic is currently in progress.

Building from the idea that red giant evolution is the main driver of the spectro-seismic connection, we can make a number of testable predictions for the observable properties of RGB versus RC stars:
\begin{itemize}
    \item The spectroscopic differentiation of upper RGB stars from RC stars will be more difficult than for lower RGB stars due to less distinct abundances. This does not present a significant difficulty, since these stars are more easily differentiated in \teff\ - \logg - luminosity space. 
    \item The spectro-seismic connection should be more pronounced in metal-poor red giants because deep mixing is much more efficient in red giants with lower metallicity. For example, \citet{Martell08} found that the carbon depletion rate from deep mixing in red giants is doubled at a metallicity of $\text{[Fe/H]}=-2.3$ as compared to $\text{Fe/H}=-1.3$.  
    \item Oxygen may also be a spectro-seismic indicator because it is also slightly depleted during phases of deep mixing \citep{Weiss00,Johnson12}. We do not find a significant correlation between atomic oxygen spectral features and the evolutionary state of red giants in our data. In addition, there are no CO lines present in the wavelength coverage of \textit{Veloce Rosso}. Other moderate- to high-resolution spectroscopic surveys that derive oxygen abundance, e.g. APOGEE \citep{Majewski17}, may be able to investigate this further.
    \item Lithium is often of high interest in studies of red giants; however, it will not be useful for this kind of study. 
    Lithium is almost completely depleted by the first dredge-up and the strong mixing at the RGB bump \citep{Lind09}, leaving little behind to allow for the detection of a significant difference between the lower RGB and RC.
    \item Previous studies \citep[e.g.,][]{Hawkins18,Ting18,Banks23} have discussed whether the helium flash may also be a driver of the spectro-seismic connection seen in red giants. Comparing the abundances of secondary RC stars to primary RC stars with the same metallicity might be a way to test this, since the secondary RC stars ignite helium fusion smoothly, without a flash.
\end{itemize}

In the pursuit of spectroscopically identifying RC stars, 
we have found that stars falling within the specific ranges of \teff, \logg\ and [Fe/H] explored in this study ($4330<T_\text{eff}<5030$\,K, $2.2<\log g<2.5$, and $-0.95<\text{[Fe/H]}<0.31$) have spectral features identifiable with high-resolution spectra that are effective for RC classification. Leveraging these features is particularly valuable for Galactic archaeology, as RC stars serve as standard candles. 
While the limiting magnitude of \textit{Veloce Rosso}, $T\sim12\,$mag, confines observations to a relatively limited volume of the Galaxy compared to those of \tess\ and \textit{Kepler} \citep[$\sim14\,$mag;][]{Hekker11,Stello22}, several other spectroscopic instruments such as VLT/UVES, Magellan/MIKE, Keck/HIRES, and Gemini/GHOST \citep[][respectively]{Dekker00,Bernstein03,Vogt94,Rantakyro24} provide comparable high-resolution capabilities to \textit{Veloce Rosso} while offering the advantage of fainter limiting magnitudes, $\sim19\,$mag.
This enables the accurate mapping of stellar population and abundance trends over a significantly larger volume of the Galaxy than \textit{Gaia} parallax. The reach of RC star samples in spectroscopic surveys will expand from the 6 kpc covered by current-generation surveys like GALAH \citep{Buder21} and APOGEE \citep{Majewski17} to 16 kpc in upcoming projects such as 4MOST \citep{deJong19} and WEAVE \citep{Dalton12}, and potentially up to 100 kpc in proposed future projects like WST \citep{Pasquini18} and MSE \citep{MSE19}.

\section*{Acknowledgements}

The authors would like to thank the anonymous reviewer for their insightful comments on the contents of this paper.

The authors would also like to thank A. R. Casey for providing an updated version of \cannon\ and guidance during the use and analysis of the program. 

KAB and SLM acknowledge funding support from the UNSW Scientia program. SLM is supported by the Australian Research Council through Discovery Project grant DP180101791. DS is supported by the Australian Research Council through Discovery Project grant DP190100666.

The authors acknowledge the Traditional Custodians of the land on which the analysis for this investigation was conducted, the Bedegal people of the Eora nation. The authors also acknowledge the Traditional Custodians of the land on which the Anglo-Australian Telescope is situated, the Gamilaraay people. The authors pay respect to elders both past and present and extend that respect to other Aboriginal and Torres Strait Islander peoples reading this paper.

\section*{Data Availability}
 
The data in this work were obtained with the \textit{Veloce} spectrograph at the Anglo-Australian Telescope through programs A/2020B/18 and A/2021B/07. Raw data are available from the AAT archive through AAO Data Central  (archives.datacentral.org.au). \tess\ asteroseismic classifications from \citet{Reyes22} can be searched and downloaded via the VizieR astronomical catalogue service.



\bibliographystyle{mnras}
\bibliography{main}




\appendix

\section{Prediction of RC/RGB Evolutionary State} \label{sec:predicting_state}

In this section, we investigate the ability of \cannon\ to accurately predict red giant evolutionary state from their spectra. While \cannon\ is not strictly a machine-learning classification algorithm, its prediction of RC and RGB labels from stellar spectra may be influenced by unbalanced training sets, as we have in this investigation, similar to other machine-learning classification methods. This is particularly prevalent in data sets with class inseparability \citep{Galar11,Hosenie20}.

To investigate potential biases present in the classifications made by \cannon, particularly those susceptible to the division between RC and RGB stars in the training set, we conduct a comparative analysis of the \texttt{ev\_state} predictions from a variety of models on unbiased test sets (i.e. balanced test sets selected from stars with known asteroseismic classifications). We compare the predictions generated by the bootstrap models from Section \ref{subsec:validation} with those from an additional set of 1,000 bootstrap models trained on a balanced distribution of 25 randomly selected RC stars and 25 randomly selected RGB stars.

We first look at the \texttt{ev\_state} predictions made in the bootstrap models from Section \ref{subsec:validation}. These \cannon\ models include training sets with a 3:1 ratio of RC to RGB stars. We evaluate the ability of \cannon\ to classify red giant stars with the following metrics: 
\begin{itemize}
    \item \textbf{Precision:} the ratio of true positives to the sum of true positives and false negatives. For RC stars, this is defined by the number of correctly predicted RC stars divided by the total number of RC predictions
    \item \textbf{Recall:} the ratio of true positives to the sum of true positives and false negatives. For RC stars, this is defined as the number of correctly predicted RC stars divided by the total number of known RC stars
    \item \textbf{$F_1$ score:} the harmonic mean of precision and recall
    \item \textbf{Accuracy:} the fraction of correct predictions out of all predictions
\end{itemize}

\begin{figure}
    \centering
    \includegraphics[width=0.481\textwidth]{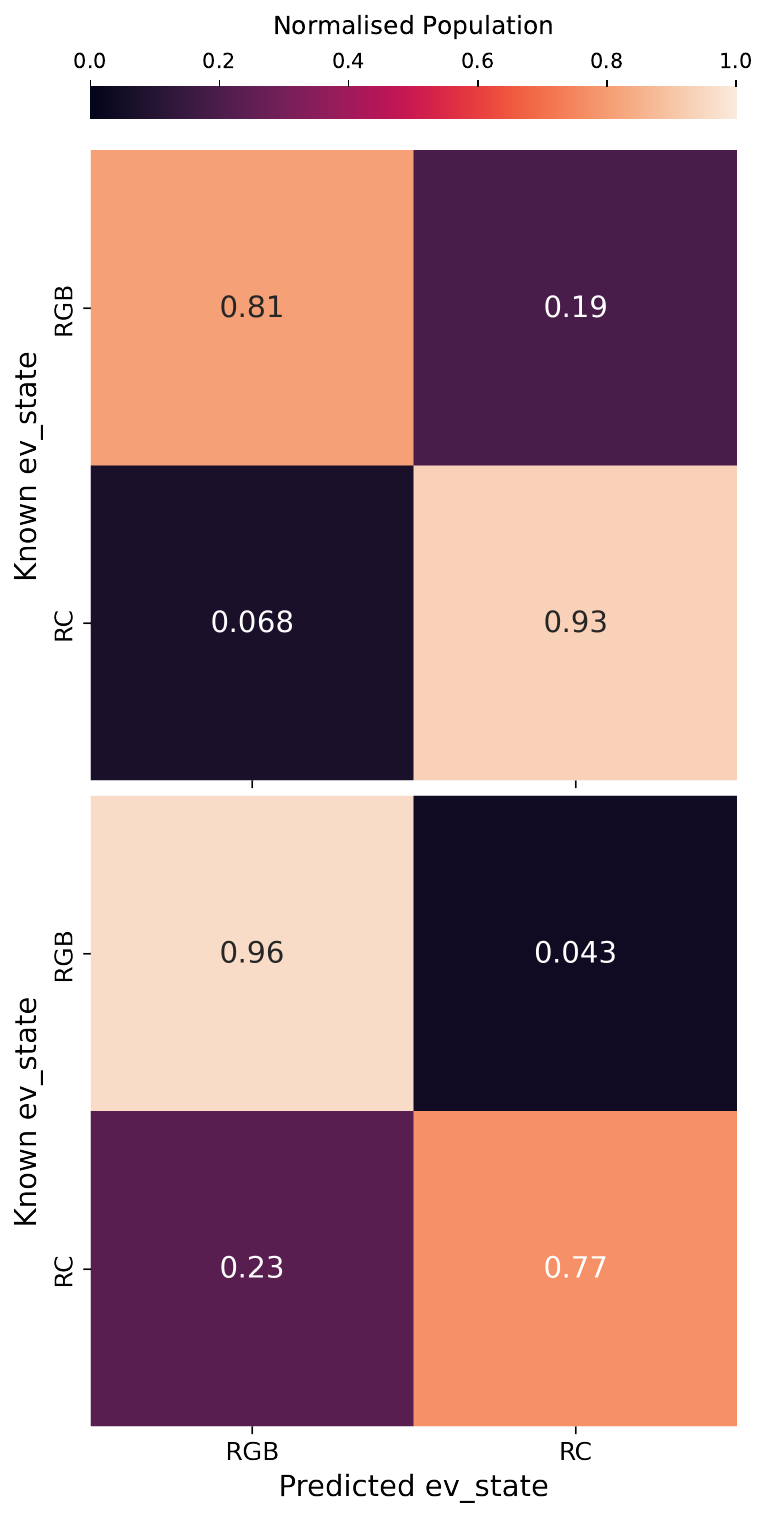}
    \caption{Confusion matrices of evolutionary state predictions made across all bootstrap models described in Section \ref{subsec:validation} with $\rm{RC:RGB}=\rm{3:1}$ (top panel) and models with $\rm{RC:RGB}=\rm{1:1}$ (bottom panel). The known evolutionary states from \tess\ asteroseismology are on the y-axes and the predicted evolutionary states made by \cannon\ are on the x-axes. True positives, i.e. RC stars correctly classified by \cannon\ as belonging to the RC occur with 93 per cent of the RC population with a 3:1 training set split and 77 per cent of the population with a 1:1 training set split.
    True negatives, i.e. RGB stars correctly classified by \cannon\ as belonging to the RGB occur with 81 per cent of the RGB population with a 3:1 training set split and 96 per cent of the population with a 1:1 training set split.}
    \label{fig:confusion_matrix}
\end{figure}

Table \ref{tab:test_scores} shows the precision, recall and $F_1$ score of all predictions made with \cannon\ in the bootstrap models with $\text{RC:RGB}=\text{3:1}$ training sets. The total accuracy of the predictions made by these \cannon\ models is 0.87. We also show the confusion matrix of these predictions in the upper panel of Figure \ref{fig:confusion_matrix}.

\begin{table}
    \centering
    \caption{Classification scores: precision, recall and $F_1$ scores for classifications made in the bootstrap models with $\rm{RC:RGB}=\rm{3:1}$ training sets. The overall accuracy of the classifications made with these models is 0.87.}
    \label{tab:test_scores}
    \begin{tabular}{cccc}
         \hline
         Evolutionary State & Precision & Recall & $F_1$ Score \\
         \hline
         RGB & 0.92 & 0.81 & 0.86 \\
         RC & 0.83 & 0.93 & 0.88 \\
         \hline
    \end{tabular}
    
    \centering
    \caption{Classification scores: precision, recall and $F_1$ scores for classifications made in the bootstrap models with training sets comprised of an equal split between RC and RGB stars. The overall accuracy of the classifications made with these models is 0.86.}
    \label{tab:test_scores_5050}
    \begin{tabular}{cccc}
         \hline
         Evolutionary State & Precision & Recall & $F_1$ Score \\
         \hline
         RGB & 0.81 & 0.96 & 0.88 \\
         RC & 0.95 & 0.77 & 0.85 \\
         \hline
    \end{tabular}
\end{table}

From these bootstrap models, true positives, i.e. RC stars accurately predicted as RC by \cannon, occur in 93 per cent of the total population of RC stars in the test sets of all models. True negatives, i.e. RGB stars accurately predicted as RGB by \cannon, occur in 81 per cent of the total population of RGB stars in the test sets of all models. Therefore, some misclassifications occur within these predictions, with RGB stars more likely to be misclassified as RC stars.

We also present the precision, recall and $F_1$ scores for the red giant evolutionary state predictions made by \cannon\ from the additional 1,000 models trained on balanced training sets in Table \ref{tab:test_scores_5050}. The total accuracy of the predictions made by these \cannon\ models is 0.86. We also show the confusion matrix of these predictions in the lower panel of Figure \ref{fig:confusion_matrix}. 

From these additional bootstrap models trained on a 1:1 split between RC and RGB stars, true positives occur in 77 per cent of the total population of RC stars in the test sets of all models. This reduction is anticipated, given the less comprehensive training set of RC spectra, comprising of 25 RC stars, as opposed to the 54 in the models featuring a 3:1 RC to RGB ratio. Conversely, true negatives have improved to 96 per cent of the total population of RGB stars in the test sets of all models. This enhancement is attributed to a more comprehensive training set of RGB stars, with 25 stars compared to 18 in the models presented in Section \ref{subsec:validation}. In these models, RC stars are more likely to be misclassified as RGB stars. 

The purpose of red giant classification is to identify RC stars such that their unique property as a standard candle can be used in studies of Galactic archaeology. From our investigation, we find that the overall classification accuracy from both sets of models yields similar results. However, both sets of models appear to be more useful for different outcomes. For example, by training \cannon\ on sets of RC and RGB stars that are representative of the total population of red giants with known evolutionary states available to train (i.e. in our case $\text{RC:RGB}=\text{3:1}$), this maximises the number of expected RC star predictions, however, there can be up to 20 per cent contamination from misclassified RGB stars. Conversely, if the goal is to accurately identify RGB stars such that there is little contamination present in RC selections, we find the balanced training set is most suitable. However, we note that this does result in fewer correct RC star classifications. 

It is important to note that the goal of \cannon\ is spectral modelling and label transfer rather than discrete classification. In our analysis, we attribute the improved classifications for each evolutionary state to a more comprehensive training set. Specifically, there is a larger selection of RC and RGB stars spanning a broader spectrum of stellar labels explored in \cannon\ models (e.g. \teff, \logg, and [Fe/H]). The models established with \cannon\ contain a total of 55,402 pixels where each pixel is described by a quadratic function of five labels (i.e. \teff, \logg, [Fe/H], \dnu\ and ev\_state) which results in 21 total parameters. Hence, larger training sets with broad coverage of label space in each evolutionary state will better constrain these factors and effectively contribute to a refined predictive capability of \cannon. 

In conclusion, when using \cannon\ as a classification tool for red giant stars, we recommend employing a sufficiently large sample of both RC and RGB stars in the training set such that each sub-set of RC and RGB stars adequately covers a more complete distribution of other stellar labels that describe their spectra (i.e. \teff, \logg, and [Fe/H]). As a result, we are confident in the spectral features we have found as significant in the prediction of red giant evolutionary state. This is because we determined these from a model generated by \cannon\ using the full suite of RC and RGB stars available for establishing and training the model (i.e. 28 RGB and 95 RC stars).

\subsection{Comparing evolutionary state predictions to other methods}
Before the rise of asteroseismology as the ``gold standard'' of red giant classification, RC stars were distinguished from the RGB either through selection cuts in photometry and surface gravity \citep[e.g. $(J-K)$ and \logg\ in][]{Williams13}, or spectroscopically on the basis of stellar parameters (\teff, \logg, [Fe/H]) and photometry, when compared to theoretical expectation from isochrones \citep[e.g.,][]{Bovy14,Sharma18}. However, these methods result in significant RGB misclassification and therefore contamination. For example, \cite{Williams13} select RC stars as those within $0.55\leq (J-K)\leq0.8$ and $1.8\leq\log g\leq3.0$ as determined from RAVE spectra \citep{Steinmetz06} and estimated that $\sim60$ per cent of stars in their RC selection actually belonged to the RGB. Selections of RC stars from isochrones provide an improved selection with less contamination from RGB stars \citep[e.g., $\sim20$ per cent in][]{Wan15}.

Next, we compare the red giant evolutionary state predictions made by our \cannon\ model to the RC/RGB classification regime explored in \cite{Martell21}. They classify red giants using two parameters, i.e., the probability for a star belonging to the RC. This probability, called \texttt{is\_redclump\_bstep}, is determined using the \textsc{bayesian stellar parameters estimator (bstep)} \citep[see][]{Sharma18}{}{}, which provides a probabilistic estimate of intrinsic stellar parameters from observed stellar properties via the use of theoretical stellar isochrones. The second parameter used to separate the RC from the RGB in \cite{Martell21} is \textit{WISE} $W_2$ absolute magnitude. They separate the RC from the RGB with the following selection:
\begin{itemize}
    \item RC stars: \textsc{bstep} RC probability $\texttt{is\_redclump\_bstep}\geq0.5$ and absolute magnitude $|W_2 +1.63|\leq0.80$
    \item RGB stars: \textsc{bstep} RC probability $\texttt{is\_redclump\_bstep}<0.5$ or absolute magnitude $|W_2 +1.63|>0.80$
\end{itemize}

We establish an additional model that is trained on all stars with asteroseismic classifications in our data set. We perform the test step of \cannon\ on the spectra of 149 stars that satisfy both selection criteria of \cite{Martell21}.

\begin{figure}
    \centering
    \includegraphics[width=0.481\textwidth]{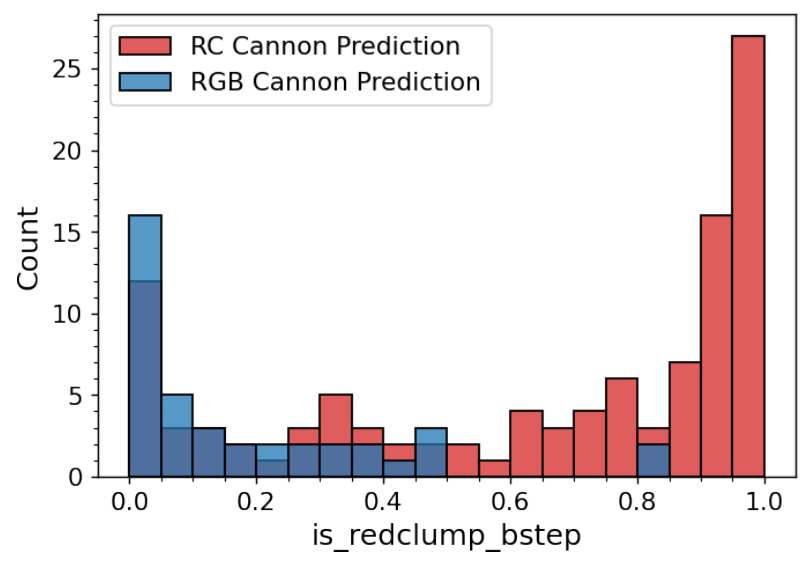}
    \caption{Results of \textsc{The Cannon} model's prediction of red giant evolutionary state compared to \texttt{is\_redclump\_bstep} score. Stars that are predicted to be from the RC by \textsc{The Cannon} are coloured in red and RGB stars in blue. The majority of stars classified by \cannon\ as belonging to the RGB agree with their \texttt{is\_redclump\_bstep} scores (where RGB stars have scores $<0.5$). However, approximately 33 per cent of stars classified as RC stars by \cannon\ also exhibit RGB \textsc{bstep} scores.}
    \label{fig:bstep}
\end{figure}

Figure \ref{fig:bstep} illustrates the evolutionary states predicted by \cannon\ compared to their \texttt{is\_redclump\_bstep} score. The red bars represent stars that are predicted to belong to the RC by \cannon\ model and the blue bars represent stars that are predicted to belong to the RGB. 

The classifications made using the scheme employed by \citet{Martell21} agree with the predictions made by \cannon\ for 98 per cent of RGB stars in our test set and 70 per cent of RC stars. To get a clear RC sample with minimal RGB contamination, from this result, we suggest using the following selection:
\begin{itemize}
    \item \cannon\ evolutionary state prediction is RC (predicted ev\_state label $\geq0.5$), and
    \item \textsc{bstep} is\_redclump\_bstep probability score is $\geq0.5$.
\end{itemize}

\section{Significant Features Line List}
\label{appendix:B}
\begin{table*}
    \centering
    \caption{Line list of the 75 spectral features within the \textit{Veloce} wavelength range found to be significant in the prediction of red giant evolutionary state and not other stellar parameters, e.g. \teff, \logg, and [Fe/H]. Here we detail the wavelength and species of the significant features, as well as their $\log gf$ and lower excitation value in eV ($E_\text{low}$) and whether this identification has changed from the investigation in \citet{Banks23} (these are identified with a checkmark).}
    \label{tab:linelist}
    \begin{tabular}{rlrrc|rlrrc}
        \hline
            Wavelength [\AA] & Species & $\log gf$ & $E_\text{low}$ [eV] & Changed ID & Wavelength [\AA] & Species & $\log gf$ & $E_\text{low}$ [eV] & Changed ID\\
        \hline
            5973.447 & $^{1}$H$^{13}$C & -4.563 & 1.971 & {} & 6495.741 & Fe I & -0.940 & 4.831 & {} \\
            5993.490 & $^{12}$C$^{14}$N & -2.566 & 0.751 & {} & 6502.418 & Ca I & -3.581 & 5.736 & {} \\
            6005.562 & $^{1}$H$^{12}$C & -4.481 & 2.666 & {} & 6512.184 & $^{12}$C$^{14}$N & -2.116 & 0.558 & {} \\
            6016.973 & Ca I & -6.553 & 6.050 & {} & 6527.634 & Mo I & -2.283 & 2.932 & {} \\
            6020.169 & Fe I & -0.270 & 4.604 & {} & 6533.955 & $^{13}$C$^{14}$N & -2.439 & 0.786 & {} \\
            6029.921 & $^{13}$C$^{14}$N & -2.429 & 0.684 & {} & 6580.208 & Ni I & -1.530 & 4.415 & {} \\
            6034.035 & Fe I & -2.420 & 4.309 & {} & 6582.882 & $^{12}$C$^{14}$N & -1.782 & 0.689 & {} \\
            6056.007 & $^{1}$H$^{13}$C & -3.535 & 1.698 & {\checkmark} & 6583.707 & Si I & -1.640 & 5.949 & {} \\
            6070.086 & $^{12}$C$^{14}$N & -2.793 & 1.880 & {} & 6595.866 & $^{1}$H$^{24}$Mg & -5.103 & 0.578 & {} \\
            6103.293 & Fe I & -1.117 & 4.730 & {} & 6611.938 & $^{12}$C$^{14}$N & -2.172 & 0.710 & {} \\
            6104.647 & $^{12}$C$^{14}$N & -1.841 & 0.975 & {} & 6617.777 & Si I & -2.380 & 6.094 & {} \\
            6129.715 & $^{12}$C$^{14}$N & -3.227 & 2.606 & {\checkmark} & 6635.731 & V II & -2.159 & 6.857 & {} \\
            6139.644 & Fe I & -4.500 & 2.586 & {} & 6668.377 & $^{12}$C$^{14}$N & -2.504 & 0.779 & {\checkmark} \\
            6169.988 & $^{12}$C$^{14}$N & -2.501 & 2.660 & {} & 6674.208 & $^{13}$C$^{14}$N & -3.017 & 3.205 & {} \\
            6217.694 & $^{1}$H$^{13}$C & -5.986 & 1.551 & {} & 6719.603 & $^{12}$C$^{14}$N & -3.075 & 2.750 & {} \\
            6225.510 & $^{1}$H$^{12}$C & -4.586 & 1.527 & {} & 6847.617 & $^{13}$C$^{14}$N & -3.946 & 3.395 & {} \\
            6243.815 & Si I & -0.770 & 5.611 & {} & 7085.484 & C I & -3.310 & 8.636 & {\checkmark} \\
            6244.501 & $^{12}$C$^{14}$N & -3.134 & 1.216 & {} & 7353.514 & $^{12}$C$^{14}$N & -3.900 & 2.610 & {} \\
            6247.559 & Fe II & -2.300 & 3.892 & {} & 7373.004 & Si I & -1.340 & 5.979 & {} \\
            6251.295 & $^{13}$C$^{14}$N & -2.980 & 1.982 & {} & 7390.289 & $^{12}$C$^{14}$N & -1.534 & 0.702 & {} \\
            6265.612 & Mn I & -1.317 & 4.229 & {} & 7405.797 & $^{12}$C$^{14}$N & -1.504 & 0.730 & {\checkmark} \\
            6272.026 & Ce II & -0.400 & 1.543 & {} & 7409.083 & Si I & -0.880 & 5.611 & {} \\
            6297.084 & $^{12}$C$^{14}$N & -3.366 & 1.585 & {} & 7423.544 & Fe I & -2.551 & 5.738 & {} \\
            6309.902 & Si I & -2.680 & 5.949 & {} & 7771.957 & $^{12}$C$^{14}$N & -3.648 & 2.278 & {\checkmark} \\
            6317.572 & $^{12}$C$^{14}$N & -2.881 & 0.190 & {} & 7894.785 & $^{13}$C$^{14}$N & -2.330 & 1.273 & {} \\
            6318.044 & $^{13}$C$^{14}$N & -3.009 & 1.247 & {} & 7905.644 & $^{12}$C$^{14}$N & -2.893 & 0.037 & {} \\
            6320.851 & Sc II & -1.770 & 1.499 & {} & 7918.375 & Fe I & -3.383 & 5.750 & {} \\
            6331.969 & Ce I & -1.803 & 0.028 & {} & 7960.716 & V I & -1.555 & 4.587 & {} \\
            6336.849 & $^{1}$H$^{13}$C & -5.386 & 1.275 & {} & 8011.986 & $^{12}$C$^{14}$N & -3.527 & 1.479 & {} \\
            6347.096 & $^{13}$C$^{14}$N & -2.668 & 1.281 & {} & 8044.402 & Fe I & -2.736 & 5.789 & {} \\
            6367.414 & $^{12}$C$^{14}$N & -3.138 & 0.266 & {} & 8071.284 & Si I & -1.420 & 6.094 & {} \\
            6371.368 & $^{12}$C$^{14}$N & -3.254 & 1.477 & {} & 8372.112 & Fe I & -1.362 & 5.788 & {} \\
            6382.641 & $^{12}$C$^{14}$N & -1.683 & 1.426 & {} & 8372.739 & $^{12}$C$^{14}$N & -1.686 & 0.592 & {} \\
            6390.498 & La II & -3.079 & 0.321 & {} & 8443.970 & Si I & -1.400 & 5.866 & {\checkmark} \\
            6407.291 & Si I & -1.500 & 5.866 & {} & 8510.285 & $^{12}$C$^{14}$N & -2.288 & 0.770 & {} \\
            6432.683 & $^{12}$C$^{14}$N & -3.153 & 2.259 & {} & 8656.674 & Fe I & -2.387 & 5.016 & {} \\
            6456.419 & $^{12}$C$^{14}$N & -2.508 & 1.476 & {\checkmark} & 8728.010 & Si I & -0.610 & 6.176 & {} \\
            6462.567 & Ca I & 0.310 & 2.521 & {} & {} & {} & {} & {} & {} \\
         \hline
            \end{tabular}
\end{table*}

\bsp	
\label{lastpage}
\end{document}